\documentclass[12pt]{article}
\usepackage{dsfont,amsthm,natbib,textcomp,amsfonts}
\usepackage[utf8]{inputenc}
\usepackage[colorlinks,citecolor=blue,linkcolor=black,urlcolor=blue]{hyperref}
\usepackage{color}

\usepackage[top=3cm, left=3cm, bottom=3cm, right=3cm]{geometry}
\usepackage{setspace}
\usepackage{graphicx}
\usepackage{bm}
\usepackage{multirow}
\usepackage{dcolumn}
\usepackage{amssymb}
\usepackage{amsbsy}
\usepackage{appendix}

\usepackage[section]{placeins}

\usepackage{subcaption}

\usepackage{sectsty}
\usepackage[normalem]{ulem}

\usepackage[multiple]{footmisc}
\usepackage{caption}

\usepackage{url}
\urldef\finraurl\url{https://www.finra.org/sites/default/files/2019%20Industry%20Snapshot.pdf}
\urldef\bankrupt\url{https://abi-org.s3.amazonaws.com/Newsroom/Bankruptcy_Statistics/Quarterlynonbusinessfilingsbychapter1994-Present.pdf}

\usepackage[figuresleft]{rotating}

\usepackage{booktabs}
\usepackage{threeparttablex}

\usepackage{siunitx}
\sisetup{
detect-mode,
group-digits            = false,
input-symbols           = ( ) [ ] - +,
table-align-text-post   = false,
input-signs             = ,
}   

\def\yyy{%
\bgroup\uccode`\~\expandafter`\string-%
\uppercase{\egroup\edef~{\noexpand\text{\llap{\textendash}\relax}}}%
\mathcode\expandafter`\string-"8000 }

\def\xxxl#1{%
\bgroup\uccode`\~\expandafter`\string#1%
\uppercase{\egroup\edef~{\noexpand\text{\noexpand\llap{\string#1}}}}%
\mathcode\expandafter`\string#1"8000 }

\def\xxxr#1{%
\bgroup\uccode`\~\expandafter`\string#1%
\uppercase{\egroup\edef~{\noexpand\text{\noexpand\rlap{\string#1}}}}%
\mathcode\expandafter`\string#1"8000 }

\def\textsymbols{\xxxl[\xxxr]\xxxl(\xxxr)\yyy}

\makeatletter
\let\estinput\@@input
\makeatother

\newcommand{\estauto}[3]{
\vspace{.75ex}{
\textsymbols
\begin{tabular}{l*{#2}{#3}}
\toprule
\estinput{#1}
\bottomrule
\addlinespace[.75ex]
\end{tabular}
}
}

\newcommand{\Figtext}[1]{%
\begin{tablenotes}[para,flushleft]
#1
\end{tablenotes}
}
\newcommand{\Fignote}[1]{\Figtext{\emph{Note:}~#1}}

\pagenumbering{arabic}

\usepackage{authblk}

\title{%
Career Incentives, Risk-Taking, and Sorting Dynamics: Evidence from Top Financial Advisers%
\footnote{%
I thank Lothar Gampel for legal assistance with data usage and Anna Ulrichshofer for research assistance. 
I also thank Mark Egan for sharing information on data construction. 
I acknowledge financial support from SFB63 and JSPS KAKENHI Grant Number JP24K04832. 
This paper was previously circulated under the title ``Career Concerns, Risk-Taking, and Upward Mobility in the Financial Services Industry: Evidence from Top Ranked Financial Advisers" (June 2020). 
}
}

\author{%
Jun Honda%
\thanks{%
Faculty of Economics and Law,
Shinshu University, Japan. %
junhonda@shinshu-u.ac.jp. %
}%
}

\date{%
March, 2025
}

\thispagestyle{empty}

\begin{document}


\thispagestyle{empty}

\maketitle

\thispagestyle{empty}

\begin{abstract}
We examine how career concerns influence the behavior and mobility of financial advisers. Drawing on a uniquely comprehensive matched panel that combines employer–employee data with a longstanding national ranking, our study tests predictions from classic career concerns models and tournament theory. Our analysis shows that, in the early stages of their careers, advisers destined for top performance differ significantly from their peers. Specifically, before being ranked, these advisers are twice as likely to obtain a key investment license, experience customer disputes at rates up to seven times higher, and transition to firms with 80\% larger total assets. Moreover, we find that top advisers mitigate the potential costs of their higher risk-taking by facing reduced labor market penalties following disciplinary actions. Leveraging exogenous variation from the staggered adoption of the Broker Protocol through an event-study framework, our results reveal dynamic sorting: firms attract high-performing advisers intensely within a short post-adoption period. These findings shed new light on the interplay between career incentives, risk-taking, and labor market outcomes in the financial services industry, with important implications for both firm performance and regulatory policy.

\bigskip



\end{abstract}

\newpage

\thispagestyle{empty}

\tableofcontents 

\thispagestyle{empty}

\cleardoublepage 

%

\pagestyle{plain} 
\setcounter{page}{1} 

\newpage


\newpage 

\section{Introduction}
\label{section: intro 0}





Understanding how career incentives shape professional behavior is central to both labor economics and organizational theory. In this paper, we investigate how financial advisers’ career concerns influence their strategic decisions
-- ranging from aggressive investments in human capital to heightened risk-taking in the face of regulatory scrutiny --
and how these decisions affect their upward mobility within the financial services industry. 
Our study is motivated by the observation that, in an industry where advisers manage large portfolios and face significant regulatory oversight, the trade-offs between signaling high productivity and incurring compliance risks are especially pronounced. This has important implications for both firm performance and regulatory policy.
Building on the foundational predictions of classic career concerns models 
\citep[e.g.,][]{fama1980agency, holmstrom1999managerial}
and insights from tournament theory 
\citep[e.g.,][]{lazear1981rank},
we hypothesize that individuals facing strong upward mobility incentives invest aggressively in credentials and adopt riskier strategies
-- interpreted in our setting as an increased tolerance for regulatory risk -- 
even if this behavior raises the likelihood of regulatory scrutiny. We test these predictions in the context of financial advisers by using compliance-related events as a proxy for risk-taking and by examining how both enhanced recruitment benefits and potential turnover costs affect career trajectories. Rather than introducing a new formal model, our empirical work leverages a comprehensive longitudinal panel that merges FINRA compliance records with national rankings from Barron’s. This rich dataset enables us to track financial advisers from their early career stages through their ascent to industry leadership, thereby examining how early aggressive actions
-- such as rapid acquisition of key licenses and heightened exposure to compliance-related events -- 
translate into enduring competitive advantages.



To rigorously test these predictions, our empirical approach combines several complementary strategies. We exploit a rich panel dataset that allows us to compare advisers within the same firm, location, and time period, thereby controlling for a wide array of unobserved factors. Moreover, by leveraging the staggered adoption of the Broker Protocol
-- a policy change that reduces job mobility frictions --
we implement 
an event-study
design to isolate the causal effects of career incentives on adviser behavior. 
This design allows us to assess whether the benefits of attracting high-performing advisers (reflected in improved upward mobility) outweigh the potential costs associated with higher turnover 
-- a trade-off that proves to be complex and, in many cases, ambiguous.
Our analysis reveals striking differences across career stages. In the early stages of their careers, top advisers are approximately 70\% more likely to acquire critical investment licenses (e.g., Series 65/66) and face up to 7 times the incidence of customer disputes compared to their peers. Although these differentials diminish over time, they remain both statistically and economically significant. Complementing these descriptive findings, our event-study analysis shows that once non-compete barriers are relaxed via the Broker Protocol, both top and average advisers experience increased mobility. Notably, top advisers are about 40\% more likely to transition to larger, Protocol-member firms in the early post-adoption period, although the net effect on firms
-- balancing enhanced recruitment against potential increases in turnover --
remains ambiguous.
Our contributions are threefold. 
First, we document how career incentives drive aggressive early-career behaviors that yield lasting competitive advantages. 
Second, by following financial advisers from entry to peak performance, we reveal heterogeneity in labor market outcomes that aggregate studies overlook. 
Third, our analysis demonstrates that institutional interventions, such as the Broker Protocol, induce dynamic sorting in the labor market, linking individual behavior with firm-level market segmentation. 
These insights not only enrich our understanding of career concerns in high-stakes environments but also have significant implications for regulatory policy and talent management.


\subsection{Related Literature}
\label{section: RL}


Our study contributes to multiple strands of literature by investigating how career incentives shape risk-taking and job mobility among financial advisers. 
Although we do not develop a new formal model, our analysis is firmly grounded in the predictions of classical career concerns frameworks
\citep[e.g.,][]{fama1980agency, holmstrom1999managerial}
and 
tournament theory 
\citep[e.g.,][]{lazear1981rank}. 
These theories suggest that individuals facing upward mobility incentives will invest aggressively in human capital and adopt riskier behavior as a means of signaling high productivity
-- even when such behavior carries potential penalties.

\paragraph{Career Concerns and Risk-Taking.}
\label{section: RL }

Empirical evidence from 
\citet{chevalier1999career}
demonstrates that mutual fund managers adjust their risk profiles in response to termination and promotion incentives. 
Similarly, 
\citet{hong2003analyzing}
document that securities analysts exhibit biased forecasts as a means of mitigating career risks. Our work extends these insights by tracking financial advisers from the outset of their careers. Unlike previous studies that focus on individuals already at high managerial levels, we examine how implicit career incentives affect risk-taking behavior
-- as measured by compliance records --
long before an adviser attains public recognition. 
This broader perspective allows us to capture the dynamics of human capital investment and risk-taking in a tournament-like environment, where high-performing advisers signal their quality while shouldering greater compliance risks.


\paragraph{Financial Adviser Misconduct.}
\label{section: RL 2}
Recent work on financial adviser misconduct has significantly advanced our understanding of how compliance records influence market outcomes. Early evidence from FINRA-based analyses 
\citep{qureshi2015investors}
highlighted that individual-level misconduct data can reveal patterns of recidivism and peer effects in compliance behavior. 
Building on these initial observations, 
\citet{egan2019market}
has provided compelling evidence
that firm-specific tolerances for compliance records lead to persistent market segmentation. 
For instance, 
they
document that some firms are more lenient toward advisers’ past disclosures, resulting in systematic differences in career trajectories across firms.

Our study extends this growing literature by focusing on a critical yet understudied segment
--
the top ranked financial advisers.%
\footnote{%
See, 
for instance,
\citet{charoenwongdoes2019,  clifford2021property, cook2020auditors, dimmock2018fraud, dimmock2018real,  gurun2021unlocking, honigsberg2021deleting, law2019financial}.
}
While previous research has primarily examined average misconduct behavior across the industry, we compare top performers with their average counterparts to uncover nuanced differences in both the frequency of compliance events and the resulting labour-market penalties. Specifically, our findings indicate that top advisers, despite experiencing a higher incidence of customer disputes, incur lower penalty rates
-- suggesting that high-performing advisers may benefit from a form of 
``labour market shielding." 
Moreover, by considering the moderating role of client composition 
(i.e., interactions with elite client types), we shed further light on how industry-specific factors can mitigate the adverse effects of misconduct.

By linking individual-level compliance behavior to labour-market outcomes, our paper contributes to the broader literature on financial misconduct and career concerns. It demonstrates that career incentives play a critical role in moderating the penalties associated with high-risk behavior
-- a dynamic that has important implications for both market segmentation and regulatory policy.

\paragraph{Job Mobility Frictions and Non-Compete Agreements.}
\label{section: RL }

A growing body of research 
\citep[e.g.,][]{clifford2021property, gurun2021unlocking}
has examined how job mobility frictions, often driven by non-compete or solicitation restrictions, affect the career trajectories of financial professionals. Our study contributes to this literature by exploiting the quasi-experimental variation provided by the Protocol for Broker Recruiting. We demonstrate that reducing mobility frictions induces dynamic sorting in the labour market: firms that join the Protocol can more effectively recruit high-performing advisers, particularly during a short, critical window. This finding not only complements existing studies but also bridges to the broader non-compete literature by showing how institutional constraints influence both individual behavior and firm-level outcomes.
In doing so, our work highlights that reducing such frictions can induce dynamic sorting in the labour market, with important implications for talent allocation and competitive firm performance.



\section{Data}
\label{section: data 0}


Our empirical analysis relies on two primary datasets: (i) the national ranking of top financial advisers compiled by Barron’s, and (ii) individual-level adviser data from the FINRA BrokerCheck database (accessed via the Central Registration Depository, CRD). We now describe these sources and the construction of our panel in detail.

\subsection{The National Ranking for Top Financial Advisers}
\label{section: data 1}


In the United States, three major rankings for financial advisers exist—Barron’s, Financial Times, and Forbes—each employing distinct criteria based on asset-under-management, industry experience, and compliance records. We focus on the Barron’s ranking for two main reasons. First, it is the oldest and longest-running ranking (initiated in 2004); second, its evolution
-- from a Top 100 list (2004–2008) to a Top 1000 list (2009–2013) and a Top 1200 list (since 2014), with an additional Top 100 Women ranking (2006–2018) -- 
renders it ideally suited for panel data analysis across states and over time. 
We manually collected and merged these annual ranking tables to classify advisers as ``top ranked" (i.e., selected at least once) versus average. 

\subsection{FINRA BrokerCheck Database}
\label{section: data 2}

The FINRA BrokerCheck database (primarily based on Form U4) provides detailed information on advisers’ employment and registration histories, professional licenses, and compliance records (e.g., customer disputes and disciplinary actions). We merge the Barron’s ranking data with FINRA records to construct an unbalanced adviser-year panel for 2000-2018. 
Although the ranking begins in 2004, extending the panel to 2000 is essential for capturing early-career behavior. 
Observations before 2000 are excluded due to data consistency issues following the transition from paper to web-based applications. 
Our final matched panel comprises approximately 12.9 million adviser-year observations on roughly 1.3 million advisers, of whom about half have exited the industry during the period.

We address survivorship bias by noting that observations from 1939 to 1999 exhibit substantial attrition. 
Following the approach of 
\citet{egan2019market} and \citet{gurun2021unlocking},
we restrict our main analysis to a ten-year window and control for cohort effects using detailed fixed effects (firm-county-year-license-experience).

\subsection{Additional Data}
\label{section: data 3}

\paragraph{Firm-Level Data.} 
To measure upward mobility, we supplement our panel data with firm-level financial information from Audit Analytics and SEC Form ADV.
Audit Analytics provides year-end revenue and total assets for broker-dealer 
(FINRA-member) firms; 
these figures are derived from the FOCUS report (SEC Rule 17a-5). 
In contrast, SEC Form ADV offers data on regulatory assets under management and the total number of client accounts for investment advisory (SEC-member) firms. 
Because Form ADV includes firm identifiers that align with our panel data, whereas Audit Analytics provides only legal firm names, we match firm names from Audit Analytics with those in the FINRA BrokerCheck and SEC Form ADV databases to assign the corresponding financial information.

\paragraph{Gender Identification.}
Since the FINRA BrokerCheck database lacks adviser gender, we impute gender using the R package gender 
\citep{mullen2018}
by matching first names with U.S. Social Security Administration data. This procedure yields gender information for approximately 95\% of advisers, with females constituting about 26\%. We include this measure as a control variable to account for potential gender differences in career outcomes.

\paragraph{The Protocol for Broker Recruiting.}
To study job mobility frictions, we augment our data with a list of firms that have joined the Protocol for Broker Recruiting. 
The directory
-- maintained by Carlie, Patchen, \& Murphy LLP --
provides legal firm names and dates of joining (or withdrawal). 
Merging this information with our panel enables us to construct a dummy variable for Protocol membership. 
In our sample, 
over 600 FINRA-member firms (approximately 7\% of all firms) are in the Protocol, employing about 37\% of advisers.

\subsection{Types of Financial Advisers}
\label{section: data 4}

We classify advisers into two groups: top ranked advisers (those selected at least once in the national ranking) and average advisers. Furthermore, we differentiate top advisers based on the timing of their ranking (before versus after being ranked). Because top advisers typically have significantly longer industry experience than average advisers, we control for experience using multi-dimensional fixed effects and conduct alternative analyses across distinct career stages.

\subsection{Measure of Risk-Taking Based on Disclosure Events}
\label{section: data 5}

Using FINRA Form U4, we categorize disclosure events into six broad types (covering customer disputes, disciplinary actions, regulatory actions, civil cases, and criminal charges). 
For our purposes, 
we focus on two categories: (i) Customer Disputes and (ii) Disciplinary Actions (by employers and regulators). 
Customer disputes are further split into those with settlement (subcategories (a) and (b)) and those without settlement (subcategories (c)–(f)).%
\footnote{%
Customer Dispute - 
(a) Settlement, 
(b) Award/Judgment,
(c) Closed/No Actions,
(d) Denied,
(e) Dismissed,
(f) Withdrawn.
See 
Appendix
\ref{appendix: def disclosure} 
for the definition of these disclosure events.
}
We denote these sets as follows:
\begin{eqnarray} \label{eq: disclosure set A1}
A_1 
\ & = & \ 
\{\text{Customer Disputes with settlement (a)--(b)}\}, 
\nonumber \\
A_2 
\ & = & \ 
\{\text{Customer Disputes without settlement (c)--(f)}\},
\nonumber \\
A 
\ & = & \ 
A_1 \cup A_2,
\label{def: customer}
\\
B 
\ & = & \ 
\{\text{Employer Disciplinary Actions (g)}\},
\nonumber \\
C 
\ & = & \ 
\{\text{Regulatory Actions (h)}\},
\nonumber \\
D 
\ & = & \ 
A_1 \cup A_2 \cup B \cup C.
\nonumber 
\end{eqnarray}
These measures serve as proxies for risk-taking, capturing the extent to which compliance events may impose reputational or labour-market penalties that influence career progression.

\subsection{Summary Statistics for Advisers}
\label{section: data 6}

Tables \ref{table: ss 1} and \ref{table: ss 2} 
summarize key characteristics and compliance outcomes for advisers.
Table \ref{table: ss 1}
highlights that top advisers typically have around 10 more years of industry experience, 5 years longer firm tenure, higher probabilities of remaining active and transitioning to new jobs, and work at firms that are, on average, twice as large. 
They are also over twice as likely to hold the Series 65/66 license and possess roughly 45\% more licenses overall. These differences suggest potential selection bias; we address this by comparing advisers with similar industry experience and controlling for detailed fixed effects. 
Table \ref{table: ss 2}
shows that, on an annual basis, top advisers are roughly 4 times more likely to receive customer disputes with settlements and 5 times more likely to receive disputes without settlements.
\subsection{Baseline Specification in the Linear Probability Model}
\label{section: data 7}

Our empirical strategy is designed to compare top ranked advisers with average advisers working in the same firm, location, and time, and with identical licensing and experience profiles. 
Let
$i=1,\dots,I$
denote advisers; 
$j=1,\dots,J$
firms; 
$l=1,\dots,L$
locations (county FIPS);
$q=1,\dots,Q$
occupational licenses (qualifications);
$t=2000,\dots,2018$
time (years). 

We estimate the following linear probability model:
\begin{eqnarray}
Y_{iqjlt}
& = &   \beta_1 \ \text{Top Adviser}_{i} \  + \ \beta_2 \ \text{Top Adviser}_{i} \times \text{After}_{it}  
\label{eq: model 0} \\
& &  \  + \ \beta_3 \  \bm{X_{it} } 
+ \ \mu_{qjlt} \  + \ \varepsilon_{iqjlt}, 
\nonumber
\end{eqnarray}
where 
the dependent variable ``$Y$" is a binary outcome (e.g., holding a license or experiencing a disclosure event),
``$\text{Top Adviser}$"
indicates whether adviser $i$ is top ranked, 
and 
``$\text{After}$"
equals one if the observation occurs after the adviser has been ranked. 
The vector 
$X$ 
includes controls for industry experience (and its square), tenure, and other characteristics.
For brevity, we denote by 
$\mu$ 
a set of multi-dimensional fixed effects (e.g., firm-county-license-time);
$\varepsilon$
is an idiosyncratic error term.
Standard errors are clustered by firm following
\citet{abadie2023should}.


\section{Main Results}
\label{section: result 0}

In this section, we document three key findings. 
First, even before receiving a formal ranking, top advisers exhibit distinct behavioral patterns—they are significantly more likely to obtain key investment licenses and to take on greater regulatory risks. 
Second, these behaviors vary systematically across career stages, with the most pronounced differences occurring in mid-career when incentives to signal competence are strongest. 
Finally, by exploiting exogenous variation in job mobility frictions through a difference-in-differences framework, we provide causal evidence that reducing these frictions induces dynamic sorting, enabling firms to recruit high-performing advisers more effectively.

We now detail these findings.

\subsection{
Patterns of
Human Capital Investment, 
Risk-Taking,
and
Upward Mobility
}
\label{section: result 1 0}
\subsubsection{%
Behavioral Differences Before and After Ranking%
}  
\label{section: result 1 1}

We begin by using our baseline specification 
(see Equation (\ref{eq: model 0})) 
to quantify differences between top advisers and their peers in terms of human capital investment, risk-taking, and upward mobility. 
In this model, the coefficient on the 
``$\text{Top Adviser}$" 
dummy captures the differential behavior of advisers who eventually become top performers, while the interaction term 
``$\text{Top Adviser} \times \text{After}$" 
isolates changes following the ranking event.


\paragraph{Human Capital Investment in Qualifications.}
Financial advisers must obtain various licenses to serve their clients. 
Our summary statistics 
(Table \ref{table: ss 1}) 
reveal that a vast majority of top advisers hold the Series 65/66 qualification
as investment adviser
-- a credential critical to their career progression
(see 
Appendix \ref{appendix: def exam}
for
the definition of qualifications/occupational licenses).

We estimate a linear probability model 
(see Equation (\ref{eq: model 0})), 
where the dependent variable is a dummy equal to one if adviser $i$ holds the Series 65/66 at time t. 
Table \ref{table: ba 1}
reports our parameter estimates under various specifications that incrementally add adviser controls (such as industry experience, tenure, and the number of other licenses) and fixed effects (including firm, location, time, license, and industry experience). 
In all specifications, the coefficient on our key independent variable
-- ``$\text{Top Adviser}$," which captures the difference between eventual top advisers and average advisers before ranking --
is positive and statistically significant. 
For example, in Column (2), the inclusion of adviser controls substantially improves the model fit, while the coefficient remains robust after adding firm-year-county fixed effects in Column (3). 
In our most comprehensive specification (Column (5)), where we control for firm-year-county-license-experience fixed effects, the coefficient is approximately 0.22. 
Given that the mean of the dependent variable is around 0.31, this result implies that, holding other factors fixed, top advisers are roughly 70\% more likely than average advisers to hold the Series 65/66 qualification prior to being ranked.

We further examine the interaction term 
``$\text{Top Adviser} \times \text{After}$," which captures any changes in this differential after the ranking event. 
Initially positive in Column (1), the coefficient for the interaction term becomes negative when controlling for fixed effects in Column (3). However, its magnitude remains small relative to that of the baseline ``$\text{Top Adviser}$" coefficient. 
This suggests that while the gap in licensing between the two groups declines after ranking, a substantial difference
-- roughly 50\% -- 
persists over time.

\paragraph{%
Risk-Taking: Incidence of Customer Disputes.
}


To capture risk-taking behavior, we focus on the set of customer disputes defined in 
(\ref{def: customer})
and construct a binary indicator, 
$\text{Disclosure}$, 
which equals one if adviser $i$ encounters any customer dispute (with or without settlement) in year $t$. 
We then estimate the linear probability model in Equation (\ref{eq: model 0}), replacing the dependent variable 
$Y$ with ``$\text{Disclosure}$", 
the annual incidence of customer disputes with and without settlements.

Panels (a) and (b) of 
Table \ref{table: ba 2}
report the results for disputes 
with settlements (and without settlements), respectively. 
Across all specifications, the coefficient on the key independent variable, “Top Adviser,” is positive and statistically significant, indicating that, prior to ranking, top advisers are considerably more likely to experience customer disputes than their peers. Although the inclusion of adviser controls and firm-year-county fixed effects does not dramatically increase , the magnitude of the “Top Adviser” coefficient diminishes as additional fixed effects—such as license and industry experience—are introduced. In our most comprehensive specification (Column 5), top advisers are roughly 2.5 times more likely than average advisers to incur disputes with settlements and 4.5 times more likely to incur disputes without settlements. These results suggest that top advisers not only expose themselves to higher regulatory risks but also tend to resolve the majority of these disputes without further escalation, thereby mitigating potential damage to their compliance records.

Next, we assess whether these risk-taking differentials change after an adviser is ranked by examining the interaction term “Top Adviser × After.” In Panel (a), the interaction coefficient is negative in the initial specifications but becomes statistically indistinguishable from zero once comprehensive fixed effects are included. In contrast, Panel (b) consistently shows a negative and significant interaction coefficient across all specifications. In the fully controlled model (Column 5), this implies that after ranking, top advisers are about 40\% less likely to incur customer disputes without settlement compared to before ranking. Nevertheless, even post-ranking, top advisers remain roughly 3.5 times more likely than their average counterparts to encounter such disputes. Overall, these findings indicate that while top advisers engage in higher risk-taking—as evidenced by their increased incidence of customer disputes—they appear to manage these risks in a way that minimizes adverse long-term consequences for their compliance records.

\subsubsection{
Patterns Across Different Career Stages
}
\label{section: result 2 0}
To capture how career concerns evolve over time, we partition our sample into 5-year windows of industry experience (up to 20 years). 
In these regressions, we re-estimate 
Equation (\ref{eq: model 0})
after 
(i) excluding the 
``$\text{Top Adviser} \times \text{After}$"
interaction
-- thus focusing solely on pre-ranking differences --
and 
(ii) restricting the sample to advisers whose industry experience falls within each 5-year window. 
Formally, we estimate:
\begin{eqnarray}
\label{eq: career 1} 
Y_{iqjlt}
& = &   \beta_1 \ \text{Top Adviser}_{i} 
\  + \ \beta_2 \  \bm{X_{it} } 
+ \ \mu_{qjlt} \  + \ \varepsilon_{iqjlt},
\end{eqnarray}
where 
$Y$ 
denotes the outcome of interest for adviser 
$i$ in 5-year window 
with 
license $q$
at 
firm $j$ 
in location 
$l$ 
and 
time $t$; 
$X$
is a vector of controls; 
and 
$\mu$
captures firm-year-county-license-experience fixed effects.

\paragraph{Human Capital Investment in Qualifications.}

Table \ref{table: career 1}
reports the estimates for the probability of holding the Series 65/66 qualification across career stages. 
In both specifications
-- (i) with adviser controls only and (ii) with additional firm-year-county-license-experience fixed effects --
the mean probability of holding the license increases with experience, reflecting the general accumulation of human capital over time. 
However, the differential between top and average advisers is most pronounced in the first 5-year window, where top advisers are more than twice as likely to obtain the Series 65/66 qualification. This gap gradually narrows over time, declining to approximately 55\% by the final window. 
These findings indicate that most top advisers establish their credentials early in their careers. To further account for this early advantage, we introduce a dummy variable for whether an adviser obtains Series 65/66 within the first two years of entry 
(``$\text{Career Start as Investment Adviser}$").%
\footnote{%
We set the 
``$\text{Career Start as Investment Adviser}$"
threshold at two years; our results are robust to alternative specifications using a range of cut-offs.
}  
Notably, this qualitative feature persists even when we restrict the sample to advisers who eventually hold the Series 65/66 at least once in their career.
\paragraph{Risk-Taking.}

Panels (a) and (b) of 
Table \ref{table: career 2}
report the estimates for the incidence of customer disputes
-- with settlements (disclosure set $A_1$) and without settlements (set $A_2$), respectively -- 
across the same career windows. 
In the first 5-year window, the coefficient on 
``$\text{Top Adviser}$"
is small and statistically indistinguishable from zero, implying no significant difference in risk-taking at the very start of their careers. 
In the second 5-year window, 
however,
the coefficient becomes positive and economically large: top advisers face a substantially higher likelihood of customer disputes, with estimates indicating that they are roughly 7.2 times more likely to incur disputes (especially in $A_2$) compared to their peers. 
In later
windows, 
the magnitude of the coefficient declines gradually, suggesting that the strong incentives for risk-taking are concentrated in the early- to mid-career stages.
Furthermore, the coefficient on the 
``$\text{Career Start as Investment Adviser}$''
dummy is positive
-- and statistically significant at the 5\% level in Panel (b) --
suggesting that advisers who secure the Series 65/66 early in their careers tend to experience more customer disputes than those who do not.

\paragraph{Upward Mobility.}

We further 
examine upward mobility by analyzing firm size using two distinct proxies: 
(i) the number of advisers at the firm and 
(ii) the total assets (in millions of US dollars) of the firm. 
For each proxy, we consider two cases. 
In specification (a), we assess firm size at the current employer at time $t$, capturing the static dimension of upward mobility. 
In specification (b), we focus on the size of the new firm at time $t+1$ following a job transition, which reflects the mobility outcome.
For 
specification (a), 
we 
modify
equation (\ref{eq: career 1}) 
by replacing 
the dependent variable
$Y_{iqjlt}$
with
``$\text{Log}(\text{Firm Size}_{iqjlt})$", 
which is
at time $t$.
For 
specification (b),
we replace 
the dependent variable 
with 
$\text{Log}(\text{Firm Size}_{iqjlt+1})$,
conditional on a firm switch.%
\footnote{%
Note: 
Observations in this model are fewer than in the full sample because we restrict the analysis to cases where advisers switch firms.%
}

Table 
\ref{table: career 3 1}
and 
\ref{table: career 3 2}
present the corresponding estimates. 
In specification (a), the coefficient on 
``$\text{Top Adviser}$" 
is positive and significant across all 5-year windows, with the differential ranging from an 80\% larger firm in the first window to over 2.5 times larger in later windows. In specification (b), which 
is limited to
job-to-job transitions, top advisers initially move to firms approximately 80\% larger; this gap declines to around 40\% in subsequent windows. 
These results indicate that early in their careers, top advisers secure positions at substantially larger firms than their peers. Although the magnitude of this advantage diminishes over time, a significant upward mobility gap persists.
\subsection{Labour Market Penalty Reduction}
\label{section: result 3 0}

Our previous analysis shows that, before ranking, top advisers are more likely to incur customer disputes while enjoying access to larger firms. To understand how they mitigate the potential costs of their high-risk behavior, we examine whether top advisers experience reduced labour market penalties associated with compliance events.


\subsubsection{Job Separation}
\label{section: result 3 1}

To assess the impact of disclosure events on job separation, we re‐estimate the baseline model 
(Equation (\ref{eq: model 0})) 
with two key modifications. First, we replace the dependent variable with a dummy indicating whether adviser $i$ leaves his or her firm at time $t+1$ 
(conditional on working at the firm at time $t$). 
Second, we introduce an indicator for whether adviser $i$ receives a disclosure in year 
$t$ and interact this variable with both the indicator for being a top adviser and the interaction 
``$\text{Top Adviser} \times \text{After}$." 
Our focus is on the subset of disclosure events drawn from 
-- that is, those events most directly linked to job separation, while excluding 
customer disputes without settlements $A_2$ 
(which rarely affect separation) 
and 
$B$ (which by construction always lead to termination).

Formally, we estimate:
\begin{eqnarray}
\label{eq: job separation 1}
\text{Separation}_{iqjlt+1} & = &   
\beta_1 \ \text{Top Adviser}_{i} 
\  + \ \beta_2 \ \text{Top Adviser}_{i} \times \text{After}_{it} 
\label{eq: model 2} \\
& &\ + \ \beta_3 \ \text{Disc}_{iqjlt} 
\nonumber \\
& & \ + \ \beta_4 \ \text{Disc}_{iqjlt} \ \times \ \text{Top Adviser}_{i}  
\nonumber \\
& & \ + \ \beta_5 \ \text{Disc}_{iqjlt} \ \times \ \text{Top Adviser}_{i} \ \times \ \text{After}_{it}  
 \nonumber \\
& & \ + \ \beta_6 \ X_{it}
  + \ \mu_{qjlt} \  + \ \varepsilon_{iqjlt}. 
\nonumber
\end{eqnarray}
Table \ref{table: js 1}
displays our estimates. 
The coefficients on ``$\text{Top Adviser}$" are consistently negative and significant, indicating that, in the absence of disclosures, top advisers are about 17\% more likely than average advisers to remain at the same firm. 
In contrast, the coefficients on ``$\text{Disclosure}$" are positive and sizeable
-- exceeding half the mean of the dependent variable --
which implies that a disclosure event raises the probability of job separation by at least 50\% for average advisers. 
Importantly, the interaction term 
``$\text{Top Adviser} \times \text{Disclosure}$" 
is negative and statistically significant. 
In our comprehensive specifications (Columns 3--5), this interaction suggests that top advisers are over 50\% less likely than their peers to exit following a disclosure event. 
When we further examine the ``$\text{Top Adviser} \times \text{After} \times \text{Disclosure}$" term, the coefficients are positive and similar in magnitude to those on ``$\text{Top Adviser} \times \text{Disclosure}$," indicating that the penalty reduction for top advisers is largely confined to the pre-ranking period.

\subsubsection{Job-to-Job Transitions}
\label{section: result 3 2}

We now shift our focus to job-to-job transitions to examine whether differences in labour-market penalties persist when advisers switch firms. 
Given the strong selection bias
-- top advisers, especially before ranking, are much more likely to secure new positions while many average advisers exit the industry -- 
we do not consider the overall job finding rate. Instead, we analyze differences in the size of the new firm to which an adviser transitions.

For this analysis, we modify equation (\ref{eq: model 2})
by replacing the dependent variable with 
the natural logarithm of new firm size as measured by the number of advisers and total assets 
(as in Section \ref{section: result 2 0} for upward mobility)
at time $t+1$, conditional on that 
adviser $i$ works for firm $j$ at time $t$ and switches to new firm $j^{\prime} (\neq j)$ at $t+1$.
This regression is estimated on a sample of advisers who switch firms 
without experiencing career interruptions or industry exit. 
Since we limit the sample to job-to-job transitions, this reduces sample size to a large extent. 
To compensate for that, we consider that ``$\text{Disclosure}$'' in the model is a dummy variable  for whether the adviser has received at least once a disclosure event in set $A_1 \cup B \cup C = D \backslash A_2$ (see equation (\ref{def: customer})), which we denote by $E$ below for simplicity.

Table \ref{table: jj 1} 
presents the estimates. 
Across all specifications, the coefficient on 
``$\text{Top Adviser}$"
is positive and significant, indicating that, prior to being ranked, top advisers tend to switch to firms that are substantially larger. For instance, controlling for fixed effects 
(Columns 3--5 for number of advisers, and Columns 8--10 for total assets), 
top advisers move to firms that are, on average, at least 50\% larger in terms of number of advisers and 2.5 times larger in terms of assets, conditional on not having received any disclosures in $E$. 
When the 
``$\text{Top Adviser} \times \text{After}$"
term is introduced, its negative coefficient suggests that, after ranking, top advisers switch to relatively smaller firms than before; nonetheless, the combined effect still implies a significant gap between top and average advisers. For example, the overall difference in new firm size is approximately 20\% (number of advisers) and about twofold (total assets) after ranking. Moreover, the coefficients on 
``$\text{Disclosure}$"
are negative and significant: in the presence of a disclosure event, average advisers switch to firms roughly 60\% smaller (by number of advisers) or 2.5 times smaller (by total assets) than in the absence of disclosures. Importantly, the positive coefficients on the interaction 
``$\text{Top Adviser} \times \text{Disclosure}$"
indicate that top advisers face a mitigated penalty relative to average advisers. However, this labour-market penalty reduction is significant only before being ranked, as evidenced by the similar magnitude of the 
``$\text{Top Adviser} \times \text{After} \times \text{Disclosure}$" 
coefficients.

In summary, while disclosure events induce severe labour-market penalties 
-- both in terms of higher job separation and reduced firm size in job transitions --
these penalties are significantly attenuated for top advisers in the pre-ranking period.

\subsection{Reduction of Job Mobility Frictions and Sorting Dynamics}
\label{section: result 4 0}

In this section, we examine how reducing job mobility frictions
-- via the Protocol for Broker Recruiting (“the Protocol”) --
affects the sorting of financial advisers across firms. The Protocol enables advisers to transfer their client accounts without restrictions, thereby lowering non-compete barriers. 
Under identical firm characteristics, advisers, particularly those managing large client bases (i.e., top advisers), prefer to work for Protocol-member firms. 
while firms join the Protocol primarily to enhance their ability to attract high-performing advisers, the policy may also inadvertently facilitate easier employee departures, thereby increasing turnover and client loss. As a result, the net effect of Protocol membership on firm performance remains ambiguous. 
Our analysis explores how these opposing forces shape labour market sorting dynamics, shedding light on whether the benefits in attracting top talent outweigh the potential costs of higher turnover.

In our sample, while only about 7\% of firms were Protocol members in 2016, nearly 44\% of advisers work for such firms. 
We exploit this variation in Protocol membership
-- using 
using an event-study framework that incorporates firm-year-location fixed effects
--
to examine whether sorting dynamics differ between top and average advisers over pre- and post-Protocol periods. 
(For consistency, our analysis is limited to the period 2001--2016 to capture job transitions before and after the Protocol and to avoid 
complications arising from recent firm withdrawals from the Protocol.%
)
\subsubsection{%
Descriptive Evidence on Job Transitions%
}
\label{section: result bp 1}


To illustrate the mechanism, 
Table \ref{table: bp 1}
reports summary statistics on job-to-job transitions across firms, differentiated by Protocol membership. 
Our findings reveal three key patterns:
\begin{enumerate}
    \item {\bf Recruitment Surge:}
    Among advisers originally employed at Protocol-member firms, there is a marked surge in transitions during the post-Protocol period, with the highest increase observed in the first 3-year window. 

    \item {\bf Top Adviser Prevalence:}
    Across both pre- and post-ranking periods, a substantially larger proportion of top advisers are employed at Protocol-member firms compared to average advisers. 

    \item {\bf Timing of Switches:}
    For advisers who transition to Protocol-member firms, the proportion of top advisers making the switch is significantly higher in the initial post-Protocol window relative to their average counterparts. 
\end{enumerate}
These 
descriptive
patterns suggest that 
the easing of non-compete constraints not only boosts  overall mobility 
but also 
intensifies the recruitment of high-performing advisers 
within a short period following the policy change.
\subsubsection{%
Estimation Results
on Sorting Dynamics%
}
\label{section: result bp 2}

Building on these descriptive insights, 
we formally assess sorting dynamics using the following event-study specification:
\begin{eqnarray}
\text{New Firm}^{q}_{iqj^{\prime}lt+1} 
\ & = &  \
\beta_1 \ \text{Top Adviser}_{i} \ + \ \beta_2 \  \text{Protocol}_{it} 
\label{eq: broker protocol model} 
 \\ 
& & \ + \ \beta_3 \ X_{it}
  + \ \mu_{qjlt} \  + \ \varepsilon_{iqjlt}. 
\nonumber
\end{eqnarray}
where 
$\text{New Firm}$
is a dummy for whether adviser 
$i$ transitions to a Protocol-member firm during the 
$q$-th 3-year window 
(with
$q=1$
representing the pre-Protocol window, and
$q =2,3,4$
representing successive post-Protocol windows).
Here, 
$\text{Top Adviser}$
is a dummy indicating whether adviser $i$ is top ranked, and
$\text{Protocol}$
is a dummy for whether the adviser’s original firm is in the Protocol at time $t$, 
and 
$\mu$
captures flexible fixed effects 
(e.g., firm-year, county, and license-experience).
We restrict the analysis to observations from 2001--2016.

Table \ref{table: bp 2} 
reports
our estimation results.
For average advisers, the coefficient on 
``$\text{Protocol}$" 
is negative in the pre-Protocol period 
(e.g., a 45\% lower likelihood of moving to a Protocol-member firm), 
but it becomes positive in the first post-Protocol window (increasing by 30\%) 
and peaks at 70\% in the second window before declining. 
This pattern indicates that, at the aggregate level, Protocol membership enhances the attractiveness of a firm in the immediate post-adoption period.

Turning to sorting by performance, d
the coefficient on 
``$\text{Top Adviser}$"
is statistically insignificant in the pre-Protocol window, 
but it turns positive and significant in the first post-Protocol window
-- suggesting that top advisers are approximately 40\% more likely than their peers to transition to a Protocol-member firm immediately 
after Protocol adoption.
In subsequent windows, this differential fades, 
indicating that competition for top talent is most intense shortly after the Protocol is introduced, while recruitment of average advisers remains elevated over a longer horizon.

Importantly, our results also indicate that the net effect on firms is ambiguous. Although Protocol membership facilitates the recruitment of both top and average advisers, it simultaneously lowers the barriers to employee departure, potentially increasing turnover. This trade-off underscores that the benefits in attracting high-performing advisers may be partly offset by the costs associated with higher employee churn, making the overall impact on firm performance unclear.
\section{Concluding Remarks}
\label{section: conclusion 0}

In this paper, we have examined the career concerns of financial advisers by comparing top performers with their average peers. Our analysis demonstrates that, at early career stages, top advisers invest more aggressively in human capital
(as measured by faster acquisition of key licenses), 
take substantially greater compliance risks 
(as reflected in a higher incidence of customer disputes), 
and secure employment at larger firms. 
Although these differences diminish over time, they persist throughout the career trajectory. 
Moreover, we provide evidence that top advisers benefit from a reduction in labour-market penalties when facing compliance-related disclosures, a benefit that is particularly pronounced before they are publicly ranked.


These findings have important implications. First, they underscore how career incentives can drive aggressive human capital investment and risk-taking strategies, even when such strategies entail significant potential costs. The observed penalty reduction for top advisers suggests that the labour market disciplines non-top performers more severely, thereby reinforcing existing hierarchies in the industry. Second, by leveraging the quasi-experimental variation induced by the Protocol for Broker Recruiting, we show that reducing job mobility frictions can intensify sorting dynamics: firms in the Protocol tend to attract top advisers more intensively
-- especially during the early post-Protocol period --
highlighting the role of institutional constraints in shaping career outcomes.


Beyond our core findings, we discuss several related issues. 

\paragraph{Top Managers and Labour Market Discipline.}
In addition to comparing top ranked advisers with their average counterparts, we extend our analysis by examining a benchmark group of top managers (e.g., CEOs and CFOs) who are also registered as financial advisers. Using firm ownership information from Form BD (for FINRA-registered firms) and Form ADV (for SEC-member firms), we construct a sample of top managers.
Our findings reveal stark contrasts: while top managers enjoy similar labour-market advantages as top advisers in the absence of disclosure events, their risk-taking behavior differs markedly across firm size categories. For instance, when firm size is segmented into small (1--149 registered representatives), mid-size (150--499), and large (500 or more), small-firm top managers are exposed to regulatory actions at a rate approximately four times higher than that of average advisers. In contrast, for mid-size and large firms, top managers exhibit considerably lower exposure. These results suggest that the mechanisms of labour-market discipline
-- and the associated penalty structures -- 
vary significantly by career stage and institutional context
(see Appendix \ref{app: top manager} for details).

\paragraph{Cost Gap in Settlement.}
Top ranked advisers manage client assets on a massive scale, and even before being ranked, they tend to oversee larger portfolios than their peers. Consistent with this expectation, our data indicate that settlement amounts for top advisers are, on average, twice as high as those for average advisers
(see Appendix \ref{app: cost} for details). 
This cost gap has dual implications: it implies a greater potential loss in consumer surplus in transactions involving top advisers, and it underscores the higher reputational and financial stakes firms face when employing high-performing individuals. In an optimal profit-maximization framework, these increased costs should be offset by the higher sales and productivity generated by top advisers.

\paragraph{Recidivism.}
Our analysis also examines whether the elevated risk-taking observed among top advisers persists among those with a prior history of misconduct. 
Drawing on the measure used by \citet{egan2019market}, 
we examine recidivism by comparing advisers with prior disclosure records. 
Our findings reveal that even among advisers with a history of misconduct, top ranked individuals remain significantly more likely to incur new customer disputes than their average counterparts (see Appendix \ref{app: recidivism} for details). 
This persistence suggests that the ``shielding" effect—whereby top advisers face reduced labour-market penalties
-- does not extend to mitigating their overall propensity for recurrent misconduct.


\vspace{.5cm}


Together, these additional analyses enrich our understanding of the broader implications of career concerns in the financial services industry. They indicate that while high-risk behavior may impose severe penalties on average advisers, top performers
-- whether as advisers or top managers --
are able to mitigate some of these costs, thereby reinforcing upward mobility and contributing to persistent market stratification. These insights have important implications for regulatory policy and talent management practices in high-stakes environments.


\newpage 


\appendix



\setcounter{figure}{0}
\renewcommand{\thefigure}{\Alph{section}.\arabic{figure}}


\setcounter{table}{0}
\renewcommand{\thetable}{\Alph{section}.\arabic{table}}





%



\newpage 


\section{%
Tables
in the Main Text%
}
\label{section: tables}


\begin{table}[htb!]
\centering
\footnotesize 
\setlength{\tabcolsep}{4pt}
\caption{%
Summary Statistics for Employment History and 
Professional Qualifications%
}
\label{table: ss 1}
\begin{threeparttable}
\estauto{table/table_ss_1.tex}{12}{c}
\Fignote{%
This table reports summary statistics based on adviser-year panel data for the period 2000--2018. 
The 
``Average Adviser" 
column includes all adviser-year observations, while the 
``Top Adviser" 
column includes observations for advisers who have been ranked as top performers (both before and after being ranked), 
conditional on their FINRA registration. 
The last column reports the results of T-tests assuming unequal variances. 
The variable 
``Remain at a Firm" 
denotes the percentage of advisers who remain with the same firm from one year to the next (excluding 2018). 
``Total Assets"
(in millions of US dollars) 
is derived from financial information in the FOCUS report 
(see Section \ref{section: data 3} (Firm-Level Data) for details)
and is adjusted for inflation using the annual Consumer Price Index (CPI) over 2000--2018. 
``New Employment"
represents the percentage of advisers who switch firms in the year following their employment at a given firm (conditional on leaving that firm by the end of the subsequent year). 
``Migration Across States/Commuting Zones/Counties"
indicates the percentage of advisers who relocate from one state (or commuting zone, or county) to another when switching firms; commuting zones are defined according to the 2000 ERS Commuting Zones provided by the United States Department of Agriculture. Definitions of licenses and qualifications are provided in 
Appendix \ref{appendix: def exam}.
}  
\end{threeparttable}
\end{table}	

\begin{table}[!htbp]
\centering
\footnotesize 
\setlength{\tabcolsep}{5pt}
\caption{Summary Statistics for 
the Incidence of Disclosure Events 
}
\label{table: ss 2}
\begin{threeparttable}
\estauto{table/table_ss_2.tex}{12}{c}
\Fignote{%
Observations are based on adviser–year panel data for the period 2000--2018. The ``Average" column includes all adviser-year observations, while the ``Top Adviser" column is restricted to observations for advisers who have been ranked as top performers (both before and after ranking), conditional on FINRA registration. 
The final column reports T-tests with unequal variances. 
Each value represents the annual incidence (in percentage points) of a given disclosure event
-- i.e., the proportion of adviser-year observations in which the event occurred at least once. 
Note that our data do not contain any instances of the disclosure category 
``Customer Dispute - Final."
}
\end{threeparttable}
\end{table}
\begin{table}[htb!]
\centering
\footnotesize 
\setlength{\tabcolsep}{4pt}
\caption{%
Differences in Qualification: Holding the Series 65/66 License%
}
\label{table: ba 1}
\begin{threeparttable}
\estauto{table/table_ba_1.tex}{12}{c}
\Fignote{%
Observations are based on adviser–year panel data for the period 2000--2018. 
The dependent variable is a dummy equal to one if adviser $i$ holds the Series 65/66 license at time $t$ 
(see Appendix \ref{appendix: def exam} for the definition). 
``Cumulative Number of Switching Firms"
denotes the total number of firm switches recorded for an adviser from entry in the industry up to the year preceding $t$; 
``Cumulative Number of Migration Across States"
is defined analogously for interstate (or cross-zone) moves. 
``License FE"
include fixed effects for the major licenses (Series 63, 6, 7, and 24), while ``Experience FE"
control for the number of years of industry experience since entry. 
All coefficients are expressed in percentage points. 
Standard errors (in parentheses) are clustered at the firm level.
\\
\text{* $p < 0.10$, ** $p < 0.05$, *** $p < 0.01$.}
}
\end{threeparttable} 
\end{table}	
\begin{table}[ht!]
\centering
\footnotesize 
\setlength{\tabcolsep}{5pt}
\caption{%
Incidence of Customer Disputes Before and After Being Ranked%
}
\label{table: ba 2}
\begin{threeparttable}
\begin{center}
{\bf (a) Customer Disputes with Settlement in Set $A_1$}
\end{center}
\estauto{table/table_ba_2_1.tex}{12}{c}


\begin{center}
{\bf (b) Customer Disputes without Settlement  in Set $A_2$}
\end{center}

\estauto{table/table_ba_2_2.tex}{12}{c}
\Fignote{%
Observations are based on adviser–year panel data for the period 2000--2018. The dependent variable is a dummy equal to one if an adviser receives a disclosure event in a given year. 
Panel (a) reports results for customer disputes with settlement (set $A_1$), 
and Panel (b) for disputes without settlement (set $A_2$); 
see Appendix \ref{appendix: def disclosure} for full definitions.
``Adviser Controls"
include dummy variables for major licenses, the number of non-major licenses, a female indicator, industry experience (in levels, without experience fixed effects), and tenure. 
Additionally, we control for the 
``Cumulative Number of Switching Firms"
(i.e., the total number of firm switches from entry until the previous year) and the 
``Cumulative Number of Migration Across States"
(i.e., the total number of inter-state or inter-zone moves from entry until the previous year). 
``License FEs"
denote fixed effects for the major licenses (Series 63, 6, 7, and 24), and 
``Experience FEs"
capture the number of years of industry experience since entry. 
All coefficients are expressed in percentage points. Standard errors (in parentheses) are clustered by firm.
\\
\text{* $p < 0.10$, ** $p < 0.05$, *** $p < 0.01$.}
}
\end{threeparttable}
\end{table}	
\begin{sidewaystable}[htb!]
\centering
\footnotesize 
\setlength{\tabcolsep}{5pt}
\caption{%
Differences in Qualification for Series 65/66 
Over 20-Year Industry Experience Across 5-Year Windows%
}
\label{table: career 1}
\begin{threeparttable}
\estauto{table/table_2_1.tex}{12}{c}	
\Fignote{%
Observations are based on adviser-year panel data for the period 2000--2018. 
The dependent variable is a dummy equal to one if an adviser holds the Series 65/66 license at time $t$ 
(see Appendix \ref{appendix: def exam} for its definition). 
``Cumulative Number of Switching Firms"
denotes the total number of firm switches from an adviser’s entry into the industry until the year preceding $t$. 
Similarly, 
``Cumulative Number of Migration Across States"
is the total number of interstate (or inter-commuting zone) moves from entry until the preceding year. 
``License FEs"
include fixed effects for the major licenses (Series 63, 6, 7, and 24), excluding other exams, and 
``Experience FEs"
control for the number of years of experience since entry. 
All coefficients are expressed in percentage points. 
Standard errors (in parentheses) are clustered by firm.
\\
\text{* $p < 0.10$, ** $p < 0.05$, *** $p < 0.01$.}
}
\end{threeparttable} 
\end{sidewaystable}
\begin{sidewaystable}[htb!]
\centering
\footnotesize 
\setlength{\tabcolsep}{5pt}
\caption{%
Incidence of Customer Disputes Over 20-Year Industry Experience Across 5-Year Windows%
}
\label{table: career 2}
\begin{threeparttable}
%
%
%
\begin{center}
{\bf (a) Customer Disputes with Settlement}
\end{center}		
\estauto{table/table_2_2_1.tex}{12}{c}
\begin{center}
{\bf (b) Customer Disputes without Settlement}
\end{center}		
\estauto{table/table_2_2_2.tex}{12}{c}
\Fignote{%
Observations are based on adviser–year panel data for the period 2000--2018. The dependent variable is a dummy equal to one if an adviser receives a disclosure event at least once in a given year. 
Panel (a) reports estimates for customer disputes with settlement (i.e., set $A_1$), and Panel (b) for customer disputes without settlement (i.e., set $A_2$); 
see Appendix \ref{appendix: def disclosure} for detailed definitions. 
``Top Adviser (Before Being Ranked)"
is a dummy variable equal to one if an adviser was classified as a top adviser prior to ranking. 
``Career Start as Investment Adviser"
is a dummy variable equal to one if an adviser obtains the Series 65/66 license within two years of entry into the industry. 
``Adviser Controls"
include dummy variables for the major licenses, the number of additional (non-major) licenses, a female indicator, industry experience (in levels, without experience fixed effects), and tenure. 
We further control for the 
``Cumulative Number of Switching Firms"
(i.e., the total number of firm switches from entry until the preceding year) and the 
``Cumulative Number of Migration Across States"
(i.e., the total number of interstate moves from entry until the preceding year). 
``License FEs"
denote fixed effects for the major licenses (Series 63, 6, 7, and 24), and 
``Experience FEs"
capture the number of years of industry experience. 
All coefficients are reported in percentage points. Standard errors (in parentheses) are clustered by firm.
\\
\text{* $p < 0.10$, ** $p < 0.05$, *** $p < 0.01$.}
}
\end{threeparttable} 
\end{sidewaystable}
\begin{sidewaystable}[htb!]
\centering
\footnotesize 
\setlength{\tabcolsep}{5pt}
\caption{%
Firm Size Over Industry Experience Across 5-Year Windows 
(Unconditional on Job Transitions)%
}
\label{table: career 3 1}
\begin{threeparttable}
\begin{center}
{\bf (a) Log(Number of Advisers)}
\end{center}
\estauto{table/table_2_3_1.tex}{12}{c}
\begin{center}
{\bf (b) 
Log(Total Assets) 
(in millions of US dollars)}
\end{center}
\estauto{table/table_2_3_2.tex}{12}{c}
\Fignote{%
Observations are based on adviser–year panel data for the period 2000--2018. 
The dependent variable is the natural logarithm of firm size, measured in two ways: 
(a) 
``Number of Advisers"
at the firm where adviser $i$ works at time $t$, 
and 
(b) 
``Total Assets (in millions of US dollars)"
of the firm at time $t$. 
``Top Adviser (Before Being Ranked)"
is a dummy variable equal to one if an adviser was classified as a top adviser before receiving a ranking. 
``Career Start as Investment Adviser"
is a dummy equal to one if an adviser obtains the Series 65/66 license within two years of industry entry.
``Adviser Controls"
include dummy variables for the major licenses, the number of additional (non-major) licenses, a female indicator, industry experience (in levels, without experience fixed effects), and tenure, as well as cumulative measures for the number of firm switches and the number of migrations across states from industry entry until the preceding year.
``License FEs"
denote fixed effects for the major licenses (Series 63, 6, 7, and 24), and 
``Experience FEs"
account for the number of years of experience since entry. 
Coefficients are reported in percentage points. Standard errors are clustered by firm.
\\
\text{* $p < 0.10$, ** $p < 0.05$, *** $p < 0.01$.}
}
\end{threeparttable}
\end{sidewaystable}

\begin{sidewaystable}[htb!]
\centering
\footnotesize 
\setlength{\tabcolsep}{7pt}
\caption{%
New Firm Size Over Industry Experience Across 5-Year Windows (Conditional on Switching Firms)
}
\label{table: career 3 2}
\begin{threeparttable}
\begin{center}
{\bf (a) Log(Number of Advisers)}
\end{center}
\estauto{table/table_2_4_1.tex}{12}{c}
\begin{center}
{\bf (b) 
Log(Total Assets) 
(in millions of US dollars)}
\end{center}
\estauto{table/table_2_4_2.tex}{12}{c}
\Fignote{%
Observations are based on adviser-year panel data for the period 2000--2018. 
The dependent variable is the natural logarithm of new firm size, measured in two ways: (a) 
``Number of Advisers" 
and 
(b) 
``Total Assets (in millions of US dollars)." 
This measure is conditional on an adviser working at firm 
$j$ 
at time $t$, 
leaving that firm by the end of year $t+1$, 
and obtaining new employment at a different firm $j$ at time $t+1$. 
``Top Adviser (Before Being Ranked)"
is a dummy variable equal to one if an adviser was classified as top adviser before receiving a ranking; 
``Career Start as Investment Adviser"
is a dummy equal to one if an adviser obtains the Series 65/66 license within two years of entering the industry. 
``Adviser Controls"
include dummy variables for the major licenses, the number of additional (non-major) licenses, a female indicator, industry experience (in levels, without experience fixed effects), and tenure, as well as cumulative measures for the number of firm switches and the number of migrations across states from entry until the year preceding $t$. 
``Original Firm FEs"
are fixed effects for the original firm $j$ at time $t$. 
``License FEs"
include fixed effects for the major licenses (Series 63, 6, 7, and 24), and 
``Experience FEs"
capture the number of years of experience since entry. 
All coefficients are reported in percentage points. Standard errors (in parentheses) are clustered by firm.
\\
\text{* $p < 0.10$, ** $p < 0.05$, *** $p < 0.01$.}
}
\end{threeparttable}
\end{sidewaystable}
\begin{table}[t!]
\centering
\footnotesize 
\setlength{\tabcolsep}{5pt}
\caption{%
Job Separation Following Disclosure Events in 
$A_1 \cup C$%
} 
\label{table: js 1}
\begin{threeparttable}
\estauto{table/table_js_1.tex}{12}{c}
\Fignote{%
Observations are based on adviser–year panel data for the period 2000–2018. The dependent variable is a dummy equal to one if an adviser leaves a firm at time $t+1$ (i.e., experiences job separation) following his/her employment at time $t$. 
In this specification, a disclosure event is defined as any occurrence in the set 
$A_1 \cup C$ at time $t$, 
where $A_1$ denotes customer disputes with settlement and $C$ denotes regulatory actions 
(see 
Appendix \ref{appendix: def disclosure} for detailed definitions).
The variable 
``Disclosure"
is a dummy that equals one if an adviser receives at least one disclosure event from 
$A_1 \cup C$ in a given year. 
``Adviser Controls"
include: a gender dummy; 
industry experience and its squared term (in levels, without experience fixed effects); tenure; 
dummies for the major licenses (Series 63, 6, 7, 65/66, and 24) (in the absence of license fixed effects); the number of additional licenses (excluding the major ones); and a dummy for whether an adviser has initiated a career as an investment adviser within two years of entry. 
All coefficients are reported in percentage points. Standard errors (in parentheses) are clustered by firm.
\\
\text{* $p < 0.10$, ** $p < 0.05$, *** $p < 0.01$.}
}
\end{threeparttable}
\end{table}
\begin{sidewaystable}[t!]
\centering
\footnotesize 
\setlength{\tabcolsep}{3pt}
\caption{%
New Firm Size 
Following Disclosure Events $E = A_1 \cup B \cup C$%
}
\label{table: jj 1}
\begin{threeparttable}
\estauto{table/table_jj_1.tex}{12}{c}
\Fignote{%
Observations are based on adviser-year panel data for the period 2000--2018. The dependent variable is the natural logarithm of firm size
-- measured either by the total number of advisers or by total assets (in millions of US dollars) --
at the new firm $j$ in year $t+1$. 
This measure is conditional on adviser $i$ working for firm $j$ at time $t$ and switching to a new firm $j^{\prime} \neq j$ at time $t+1$ 
(i.e., new employment). 
A disclosure event is defined as any occurrence in the set 
$E \ = \ A_1 \cup B \cup C$ at time $t$, 
where 
$A_1$ denotes customer disputes with settlement, 
$B$ denotes employer disciplinary actions (e.g., termination), and 
$C$ denotes regulatory actions 
(see Appendix \ref{appendix: def disclosure} for detailed definitions). 
The variable 
``Disclosure"
is a dummy equal to one if an adviser receives at least one disclosure event from 
$E$ in a given year. 
``Firm FEs"
denote fixed effects for the original firm $j$ at time $t$. 
``License FEs"
include fixed effects for the major licenses (Series 63, 6, 7, 65/66, and 24), and 
``Experience FEs"
capture the number of years of experience since entry into the industry. 
``Adviser Controls"
include a gender dummy; industry experience and its squared term (in levels, without experience FEs); tenure; dummies for the major licenses (without license FEs); the number of additional (non-major) licenses; and a dummy indicating whether the adviser began his/her career as an investment adviser (i.e., obtained Series 65/66) within two years of entry. 
All coefficients are reported in percentage points. 
Standard errors (in parentheses) are clustered by firm.
\\
\text{* $p < 0.10$, ** $p < 0.05$, *** $p < 0.01$.}
}
\end{threeparttable}
\end{sidewaystable}	
\begin{table}[htbp!]
\centering
\scriptsize 
\setlength{\tabcolsep}{10pt}
\caption{%
Summary Statistics for the Protocol for Broker Recruiting%
}
\label{table: bp 1}
\begin{threeparttable}
\estauto{table/table_bp_1.tex}{12}{c}	
\Fignote{
Observations are based on adviser–year panel data for the period 2004--2016. 
The sample is restricted to cases where advisers work for an 
``Original Firm"
in a given year and switch to a different 
``New Firm" 
in the subsequent year. 
We classify observations into two cases based on whether the original firm is a Protocol member and further differentiate these by whether the new firm is a Protocol member. Firm size is measured using both the number of advisers and financial metrics 
-- including Assets Under Management (AUM), Revenue, and the Total Number of Client Accounts 
(see 
Section \ref{section: data 3} 
for detailed data sources).
}
\end{threeparttable}
\end{table}
\begin{sidewaystable}[hbtp!]
\centering
\scriptsize 
\setlength{\tabcolsep}{3pt}
\caption{%
Sorting Dynamics After Firms Join the Protocol for Broker Recruiting%
}
\label{table: bp 2}
\begin{threeparttable}
\estauto{table/table_bp_2.tex}{20}{c}
\Fignote{%
Observations are based on adviser-year panel data for the period 2001--2016. 
The dependent variable is a dummy equal to one if an adviser transitions to a 
Protocol-member firm at time $t+1$, 
conditional on the adviser working for a firm in year $t$ 
and switching to a new firm by the end of year $t+1$. 
Observations of advisers who exit the industry are excluded. 
The variable 
``Protocol"
is a dummy equal to one if the original firm at time $t$ 
is in the Protocol. 
Standard errors (in parentheses) are clustered by firm.
\\
\text{* $p < 0.10$, ** $p < 0.05$, *** $p < 0.01$.}
}
\end{threeparttable}
\end{sidewaystable}

\newpage 




\section{%
Details of the Barron’s Ranking Tables for Top Financial Advisers%
}
\label{app: barrons}

The Barron's annual ranking in our sample contains the list of top ranked advisers across states over the period 2004--2018.
We collect three types of the annual ranking: 
(i) top 100 over the period 2004--2008;
(ii) top 1000 (1200) over the period 2009--2013 (2014--2018); 
(iii) top 100 women over the period 2006--2018.%
\footnote{Note that Top 100 and 1000/1200 contain both female and male advisers, although the majority of them are male.
Note also that the top 100 (top 100 Women) ranking table in the year 2005 (2008) is not available online through Barron's subscription. 
To complement this, we use the last year's ranks in the top 100 (top 100 Women) table in the subsequent year 2006 (2009), which covers the majority (roughly 70\%) of listed advisers in 2005 (2008). 
We believe that the  missing part does not affect the main result. 
}
Besides the list of ranked advisers with their affiliated firms and branch office locations, 
the annual ranking table also provides
information on the type of clients as well as the size of client assets: 
(i) (team-based) total assets and 
(ii) the average client account and (high- or ultra-high-)net-worth-client assets.
Below we will limit attention to the sample over the period 2009--2018 for simplicity and describe each of them in detail.



\subsection{Type of Clients}
\label{app: barrons 1}




The annual Barron’s ranking lists top financial advisers across states from 2004--2018. The tables provide, among other details, the type of clients served. We classify clients into six categories:
\begin{enumerate}
    \item Individuals with assets below \$1 million
    \item High-net-worth individuals with assets between \$1 and \$10 million
    \item Ultra-high-net-worth individuals with assets above \$10 million
    \item Endowments
    \item Foundations
    \item Institutions
\end{enumerate}
Table \ref{}
presents the percentage breakdown by client type for top ranked advisers. 
Our analysis shows that the vast majority serve individual (high-net-worth) clients, while a distinct subset caters to institutional, endowment, and foundation clients. This heterogeneity contributes to significant variation in client asset sizes across advisers.

\subsection{Size of Client Assets}
\label{app: barrons 2}
The Barron’s ranking varies over time: a Top 100 list was used during 2004--2008, a Top 1000 list during 2009--2013, and a Top 1200 list from 2014–2018. These differences necessitate separate analyses for the periods 2009--2013 and 2014--2018.

Table A.2 reports summary statistics for client asset sizes among top ranked advisers for 2009–2013, while Table A.3 does so for 2014–2018. In both periods, the average client asset size exceeds \$1 billion, with substantial heterogeneity across subgroups (e.g., Top 100 Women vs. overall top advisers).

\subsection{Size of Assets over Time}
\label{app: barrons 3}
Figure \ref{}
illustrates the evolution of team-based total assets over time. Notably, the average total assets of top ranked advisers approximately doubled between 2009 and 2018. In 2018, the aggregate total assets managed by all Top 1200 advisers reached approximately \$2.7 trillion. In contrast, the average client account size and high-net-worth client assets have remained relatively stable over this period.

\subsection{%
Market Size Measured by Number of Ranked Advisers Across States Before and After 2014%
}
\label{app: barrons 4}
We examine the distribution of top ranked advisers across states by comparing data from 2009--2013 and 2014--2018. Table A.4 presents the number of ranked advisers by state and introduces categorical variables (Market-Size 1 to 4) based on the number of ranked advisers. For example, Market-Size 1 comprises states with between 90 and 120 ranked advisers, Market-Size 2 includes states with 30 to 90, Market-Size 3 with 20 to 30, and Market-Size 4 with fewer than 20. These market-size categories help us assess how the allocation of top advisers differs across states relative to the total number of financial advisers.

\subsection{Relationship Between the Type of Clients and Size of Total Assets}
\label{app: barrons 5}



Using the baseline model from the main text (with the natural logarithm of total assets as the dependent variable), we examine how client type correlates with firm size among top ranked advisers. We include dummies for each client type as defined in Section A.1.

Table \ref{}
reports parameter estimates showing that advisers with predominantly small-asset clients tend to manage smaller total assets, whereas those serving high-net-worth or ultra-high-net-worth clients manage substantially larger asset pools.


\subsection{Relationship Between Ranks and Size of Total Assets}
\label{app: barrons 6}



We hypothesize that higher-ranked advisers manage larger total assets. Using the same baseline model as in A.5 and including an additional independent variable, “Rank” (representing the adviser’s current rank within a state), we test this hypothesis.

Table \ref{}
presents results across several specifications. In the absence of adviser controls, the sign on “Rank” is ambiguous; however, after including adviser and firm–year–state fixed effects, the coefficient turns negative, consistent with the expectation that a lower rank (i.e., better performance) is associated with larger total assets.

\subsection{Relationship Between Ranks and Incidence of Disclosure Events}
\label{app: barrons 7}
We further explore the relationship between an adviser’s rank and the incidence of disclosure events. We group top ranked advisers into three categories based on their state-level ranking (Rank 1: top 25\%, Rank 2: 25--50\%, Rank 3: 50--75\%, with Rank 4 as the base category for advisers with rankings above 75\%).

Table \ref{}
summarizes summary statistics for disclosure incidence across these ranking groups. Our analysis reveals that, before 2014, the top tier of ranked advisers (Rank 1) are more likely to receive customer disputes with settlement than their lower-ranked counterparts; after 2014, this pattern shifts somewhat, with the second tier (Rank 2) experiencing higher incidence. Overall, our findings indicate significant heterogeneity in disclosure events even among top ranked advisers.



\section{Definition of the Major Disclosure Events}
\label{appendix: def disclosure}


Disclosure events details are described in Form U4.%
\footnote{%
The Form U4 is available via \url{https://www.finra.org/sites/default/files/form-u4.pdf} 
(accessed February 20, 2025).
Note that the definition of each event is given in the FINRA's BrokerCheck report for financial advisers (registered representatives) who have indeed received that disclosure in the past. 
See 
\url{https://brokercheck.finra.org/} 
and also \url{https://www.finra.org/sites/default/files/AppSupportDoc/p015111.pdf}
(accessed February 20, 2025).
} 
Below we consider the major disclosure events 
excluding 
those on appeal and pending ones,
and give their definitions used in the FINRA's BrokerCheck database.%


{\footnotesize

\vspace{0.5cm} \noindent {\bf Customer Dispute - Settled:}
This type of disclosure event involves a consumer-initiated, investment-related complaint, arbitration proceeding or civil
suit containing allegations of sale practice violations against the broker that resulted in a monetary settlement to the
customer.

\vspace{0.5cm} \noindent {\bf Customer Dispute - Award / Judgment:}
This type of disclosure event involves a final, consumer-initiated, investment-related arbitration or civil suit containing
allegations of sales practice violations against the broker that resulted in an arbitration award or civil judgment for the
customer.

\vspace{0.5cm} \noindent {\bf Customer Dispute - Closed-No Action / Withdrawn / Dismissed / Denied:}
This type of disclosure event involves (1) a consumer-initiated, investment-related arbitration or civil suit containing
allegations of sales practice violations against the individual broker that was dismissed, withdrawn, or denied; or (2) a
consumer-initiated, investment-related written complaint containing allegations that the broker engaged in sales practice
violations resulting in compensatory damages of at least \$5,000, forgery, theft, or misappropriation, or conversion of funds
or securities, which was closed without action, withdrawn, or denied.

%
%
%
%

\vspace{0.5cm} \noindent {\bf Criminal - Final Disposition:}
This type of disclosure event involves a conviction or guilty plea for any felony or certain misdemeanor offenses, including
bribery, perjury, forgery, counterfeiting, extortion, fraud, and wrongful taking of property that is currently on appeal.

\noindent
{\bf Type:} Felony, Misdemeanor.

%
%
%
%

\vspace{0.5cm} \noindent {\bf Civil - Final:}
This type of disclosure event involves (1) an injunction issued by a court in connection with investment-related activity, (2)
a finding by a court of a violation of any investment-related statute or regulation, or (3) an action brought by a state or
foreign financial regulatory authority that is dismissed by a court pursuant to a settlement agreement.

%
%
%
%
%

\vspace{0.5cm} \noindent {\bf Employment Separation After Allegations:}
This type of disclosure event involves a situation where the broker voluntarily resigned, was discharged, or was permitted
to resign after being accused of (1) violating investment-related statutes, regulations, rules or industry standards of
conduct; (2) fraud or the wrongful taking of property; or (3) failure to supervise in connection with investment-related
statutes, regulations, rules, or industry standards of conduct.

\noindent
{\bf Termination Type:} Discharged, Permitted to Resign, Voluntary Resignation.

\vspace{0.5cm} \noindent {\bf Regulatory Final:}
This type of disclosure event may involve (1) a final, formal proceeding initiated by a regulatory authority (e.g., a state
securities agency, self-regulatory organization, federal regulatory such as the Securities and Exchange Commission,
foreign financial regulatory body) for a violation of investment-related rules or regulations; or (2) a revocation or
suspension of a broker's authority to act as an attorney, accountant, or federal contractor.

%
%
%
%
%
%
%

\vspace{0.5cm} \noindent {\bf Financial - Final:}
This type of disclosure event involves a bankruptcy, compromise with one or more creditors, or Securities Investor
Protection Corporation liquidation involving the broker or an organization/brokerage firm the broker controlled that
occurred within the last 10 years.

\noindent
{\bf Action Type:} 
Bankruptcy [Chapter 7, Chapter 11, Chapter 13, Other],
Compromise, Declaration, Liquidation, Receivership, Other.

\noindent
{\bf Disposition Type:} 
Direct Payment Procedure, Discharged, Dismissed, Dissolved, SIPA Trustee Appointed, Satisfied/Released, Other.

%
%
%
%
%

\vspace{0.5cm} \noindent {\bf Judgment / Lien:}
This type of disclosure event involves an unsatisfied and outstanding judgments or liens against the broker.

\noindent
{\bf Type:} Civil, Tax.

\vspace{0.5cm} \noindent {\bf Civil Bond:}
This type of disclosure event involves a civil bond for the broker that has been denied, paid, or revoked by a bonding
company.

\vspace{0.5cm} \noindent {\bf Investigation:}
This type of disclosure event involves any ongoing formal investigation by an entity such as a grand jury state or federal
agency, self-regulatory organization or foreign regulatory authority. Subpoenas, preliminary or routine regulatory inquiries,
and general requests by a regulatory entity for information are not considered investigations and therefore are not
included in a BrokerCheck report.

}



\newpage


\section{Definition of the Major Qualification Exams (Licenses)}
\label{appendix: def exam}

The definitions of qualification exams (licenses) are described in the FINRA website.%
\footnote{See the website: 
\url{https://www.finra.org/registration-exams-ce/qualification-exams}
(accessed February 20, 2025).
} 
Below we consider the major qualification exams (Series 6, 7, 24, 63, 65, 66) as in the main text and give their definitions used in the website.
Series 6 and 7 are categorized as ``FINRA Representative-level Exams", 
Series 24 as ``FINRA Principal-level Exams",  
Series 63, 65, and 66 as ``North American Securities Administrators Association (NASAA) Exams". 
Note that the definitions of NASAA Exams are given by the NASAA website.%
\footnote{See the website: 
\url{https://www.nasaa.org/exams/exam-study-guides/}
(accessed February 20, 2025).
}

{\footnotesize

\vspace{0.5cm} \noindent {\bf Series 6:}
The Series 6 exam -- the Investment Company and Variable Contracts Products Representative Qualification Examination (IR) -- assesses the competency of an entry-level representative to perform their job as an investment company and variable contracts products representative.
The exam measures the degree to which each candidate possesses the knowledge needed to perform the critical functions of an investment company and variable contract products representative, including sales of mutual funds and variable annuities.

\vspace{0.5cm} \noindent {\bf Series 7:}
The Series 7 exam -- the General Securities Representative Qualification Examination (GS) -- assesses the competency of an entry-level registered representative to perform their job as a general securities representative. 
The exam measures the degree to which each candidate possesses the knowledge needed to perform the critical functions of a general securities representative, including sales of corporate securities, municipal securities, investment company securities, variable annuities, direct participation programs, options and government securities.

\vspace{0.5cm} \noindent {\bf Series 24:}
The Series 24 exam -- the General Securities Principal Qualification Exam (GP) -- assesses the competency of an entry-level principal to perform their job as a principal dependent on their corequisite registrations.
The exam measures the degree to which each candidate possesses the knowledge needed to perform the critical functions of a principal, including the rules and statutory provisions applicable to the supervisory management of a general securities broker-dealer.%
\footnote{In addition to the Series 24 exam, candidates must pass the Securities Industry Essentials (SIE) Exam (since October 1, 2018 with a complete overhaul) and a representative-level qualification exam, or the Supervisory Analysts Exam (Series 16) exam, to hold an appropriate principal registration. 
See the FINRA website for the definitions of related exams. }

\vspace{0.5cm} \noindent {\bf Series 63:}
The Series 63 exam -- the Uniform Securities State Law Examination -- is a North American Securities Administrators Association (NASAA) exam administered by FINRA.


\noindent	
(Definition given by NASAA:)
The Uniform Securities Agent State Law Examination was developed by NASAA in cooperation with representatives of the securities industry and industry associations. The examination, called the Series 63 exam, is designed to qualify candidates as securities agents. The examination covers the principles of state securities regulation reflected in the Uniform Securities Act (with the amendments adopted by NASAA and rules prohibiting dishonest and unethical business practices). The examination is intended to provide a basis for state securities administrators to determine an applicant?s knowledge and understanding of state law and regulations.

\vspace{0.5cm} \noindent {\bf Series 65:}
The Series 65 exam -- the NASAA Investment Advisers Law Examination -- is a North American Securities Administrators Association (NASAA) exam administered by FINRA.


\noindent	
(Definition given by NASAA:)
The Uniform Investment Adviser Law Examination and the available study outline were developed by NASAA. The examination, called the Series 65 exam, is designed to qualify candidates as investment adviser representatives. The exam covers topics that have been determined to be necessary to understand in order to provide investment advice to clients.

\vspace{0.5cm} \noindent {\bf Series 66:}
The Series 66 exam -- the NASAA Uniform Combined State Law Examination -- is a North American Securities Administrators Association (NASAA) exam administered by FINRA.


\noindent	
(Definition given by NASAA:)
The Uniform Combined State Law Examination was developed by NASAA based on industry requests. The examination (also called the ``Series 66") is designed to qualify candidates as both securities agents and investment adviser representatives. The exam covers topics that have been determined to be necessary to provide investment advice and effect securities transactions for clients.%
\footnote{%
The FINRA Series 7 is a corequisite exam that needs to be successfully completed in addition to the Series 66 exam before a candidate can apply to register with a state.
}

}

\section{%
Top Managers}
\label{app: top manager}

\section{%
Cost Gap in Settlement%
} 
\label{app: cost}

We have seen 
in the main text
that top advisers are way more likely than average to encounter customer disputes. 
We now take a look at the associated cost, which can significantly differ between top advisers and average ones due to a number of possible reasons. 
One of them would be that top ranked advisers deal with (i) a larger set of clients, (ii) high net-worth clients more often, and (iii) a wider range of products and services than average advisers, even before being ranked.
As such, we expect that settlement associated with customer disputes 
are significantly higher for top advisers, which will be shown below.

Since the associated cost with disclosures is of interest, we pay attention to the set of disclosure events in which top advisers reached  pay settlements, so that we can examine a potential pay gap between themselves and average advisers. 
To illustrate
this, 
we first take a look at 
(i) the fraction of disclosure events with (pay) settlement
and 
(ii) the associated monetary compensation amount.%
\footnote{The amount is adjusted with Consumer Price Index (CPI) by the US Bureau of Labor Statistics. }
Table 
\ref{appendix table: settlement summary statistics} 
provides 
these for both top ranked advisers and average ones, 
and shows
that 
the majority of 
pay settlements 
stem from
a single disclosure event ``{\footnotesize Customer Dispute - Settled}" (see Section \ref{appendix sec: definition disclosure events} for the definition). 
We can also see that 
there are differences between top and average advisers:
(i) 
The monetary amount is way larger for top advisers than for average ones;
(ii) 
The fraction of pay settlements attributed to ``{\footnotesize Customer Dispute - Settled}"
is roughly 90\% 
for top advisers 
while around 55\% for average advisers.
Below we restrict our attention to the set of ``{\footnotesize Customer Disputes with Settlement}'' defined by $A_1$ in 
the main text. 

To examine the difference, we consider the baseline model 
in the main text
with replacing the dependent variable by the (natural) logarithm of the money amount paid in settlement. 
Table 
\ref{table: settlement customer settled details} 
provides the parameter estimates across different specifications. 
Note that unlike in the main analysis, we consider firm-year-county-license fixed effects without including industry experience fixed effects, as there are insufficient observations, leading to noisy estimates. 

We first look at the coefficient of ``$\text{\footnotesize Top Adviser}$", which is positive and significant at any reasonable level.
The coefficient in column (8)  is equal to roughly 0.7, 
which indicates that the money amount in settlements increases roughly twice
for top ranked advisers, compared to average advisers who work for the same firm, at the same time, and at the same location.%
\footnote{The coefficient 0.7 is evaluated at $100 \times (\exp(0.7)-1) \approx 100\%$. } 
To put this in perspective, the mean of the dependent variable in column (6) is around 
$10.85$, 
which indicates that 
a settlement costs over 
$\$50,000$ 
on average, whereas this amount increases to over 
$\$100,000$ 
in the case of top advisers.

Next, we take a look at the coefficient of ``$\text{\footnotesize Top Adviser} \times \text{\footnotesize After}$". 
Despite being statistically insignificant, it is negative and suggests that the money amount for top advisers decreases to some extent after being ranked but there is a persistent difference between 
the
two groups. 

Overall, 
we have seen that before being ranked, the money amount in settlements for top ranked advisers is approximately twice as large as for average ones. 	


Top ranked advisers manage larger client assets, which we expect to result in higher settlement amounts in customer disputes. In this section, we examine two aspects:

The fraction of disclosure events resulting in pay settlements
The monetary amount of settlements.
Table \ref{} 
reports that the vast majority (approximately 90\%) of pay settlements for top advisers stem from customer disputes with settlement, compared to about 55\% for average advisers. Using the baseline model with the natural logarithm of settlement amounts as the dependent variable, 
Table \ref{} 
shows that the coefficient on ``Top Adviser" is around 0.7, implying that settlement amounts for top advisers are roughly twice as high as for average advisers. For context, the mean settlement amount in the baseline specification is approximately 10.85 (in log terms), indicating that, on average, settlements exceed \$50,000, with top advisers’ settlements exceeding \$100,000.


\section{%
Prior Records and Recidivism%
} 
\label{app: recidivism}
We explore whether top advisers continue to exhibit higher levels of risk-taking (in terms of receiving customer disputes) even among those with prior misconduct records. Following the approach of Egan, Matvos, and Seru (2016), we introduce the variable “Prior Disclosure,” a dummy equal to one if an adviser has received a disclosure in set 
 before time 
. We then re-estimate the baseline model with the interaction terms “Top Adviser × Prior Disclosure” and “Top Adviser × After × Prior Disclosure” to capture differences in recidivism between top and average advisers.

Tables D.1 and D.2 present the results for customer disputes with settlement (set 
) and without settlement (set 
), respectively. Our estimates indicate that, conditional on prior disclosures, advisers are approximately 6 times (for 
) and 5 times (for 
 ) more likely to incur new disputes compared to advisers without prior records. Furthermore, even after accounting for prior records, the “Top Adviser” coefficient remains above 2 (for 
) and above 4 (for 
), suggesting that top advisers are still significantly more likely than their peers to incur new disclosures. The interaction terms confirm that this effect is most pronounced before ranking, while the differential diminishes post-ranking.



\newpage 

\bibliographystyle{ier}
\bibliography{reference/bibtex_ia} 

\end{document}